\documentclass[]{aastex}
\usepackage{emulateapj5}

\usepackage{graphicx,graphics,epsfig,amssymb,epsf}
\newcommand{\ki}{$\chi^2$ }
\newcommand{\eg}{{\it e.g.}}
\newcommand{\etal}{et~al.~}

\newcommand{\kir}{$\chi^2_{\rm red}$}

\newcommand{\sw}{SWIFT~J1753.5$-$0127}
\submitted{Accepted in ApJ}
\shorttitle{Multiwavelength observations of SWIFT~J1753.5$-$0127}
\shortauthors{Cadolle Bel et al.}

\begin{document}

\title{Simultaneous multiwavelength observations of the Low/Hard State of the 
X-ray transient source SWIFT~J1753.5$-$0127}

\author{M. Cadolle Bel\altaffilmark{1,2,3}, M. Rib\'o\altaffilmark{1,4,5}, 
J. Rodriguez\altaffilmark{1,4}, S. Chaty\altaffilmark{1,4}, 
S. Corbel\altaffilmark{1,4}, A. Goldwurm\altaffilmark{1,2}, 
F. Frontera\altaffilmark{6}, R. Farinelli\altaffilmark{6}, P. D'Avanzo\altaffilmark{7}, 
A. Tarana\altaffilmark{8}, P. Ubertini\altaffilmark{8},
P. Laurent\altaffilmark{1,2}, P.~Goldoni\altaffilmark{1,2} and 
I. F. Mirabel\altaffilmark{9} }

\altaffiltext{1}{\scriptsize Service d'Astrophysique, 
DSM/DAPNIA/SAp, CEA-Saclay, 91191 Gif-sur-Yvette, France}

\altaffiltext{2}{\scriptsize AstroParticule et Cosmologie, UMR 7164, 
(CEA/CNRS/Universit\'e Paris 7/Observatoire de Paris), France}

\altaffiltext{3}{\scriptsize European Space Astronomy Centre (ESAC), 
Madrid (Spain); Marion.Cadolle@sciops.esa.int}

\altaffiltext{4}{\scriptsize Astrophysique Int\'eractions Multi-\'echelles,
UMR 7158 (CEA/CNRS/Universit\'e Paris 7), France}

\altaffiltext{5}{\scriptsize Universitat de Barcelona, Spain}

\altaffiltext{6}{\scriptsize TESRE/CNR, Bologna \& Universit\`a 
di Ferrara, Italy}

\altaffiltext{7}{\scriptsize INAF, Osservatorio Astronomico di Brera, 
Merate \& Universit\`a degli Studi dell'Insubria, Como, Italy}

\altaffiltext{8}{\scriptsize IASF/INAF Roma \& Universit\`a di Roma 
Tor Vergata, Italy}

\altaffiltext{9}{\scriptsize ESO, Santiago, Chile}

\begin{abstract}

We report the results of simultaneous multiwavelength observations of the
X-ray transient source \sw~performed with {\it INTEGRAL}, {\it RXTE}, 
NTT, REM and VLA on
2005 August 10--12. The source, which underwent an 
X-ray outburst since 2005 May
30, was observed during the {\it INTEGRAL} Target of Opportunity program
dedicated to new X-ray novae located in the Galactic Halo. Broad-band spectra
and fast timing variability properties of \sw~are analyzed together with 
the optical, near infra-red and radio data. 
We show that the source 
was significantly detected
up to 600~keV with Comptonization parameters and timing properties typical of
the so-called Low/Hard State of black hole candidates. 
We build a spectral energy distribution and we 
show that \sw~does not follow the usual radio/X-ray
correlation of X-ray binaries in the Low/Hard State.
We give estimates of distance and mass. We conclude that 
\sw~belongs to the X-ray nova class and that it is likely a black
hole candidate transient source of the Galactic Halo which 
remained in the Low/Hard State during its main outburst. We 
discuss our results within the context of Comptonization 
and jet models.

\end{abstract}

\keywords{black hole physics -- stars: individual: SWIFT~J1753.5$-$0127 
-- gamma rays: observations -- X-rays: binaries
-- X-rays: general -- optical: general -- infra-red: binaries -- 
radio continuum: general}

\section{Introduction}

X-ray Novae (XN), also known as Soft X-ray 
Transients, are Low-Mass accreting
X-ray Binaries (LMXB) usually in quiescent state. They may undergo sudden and
bright few-month-long X-ray outbursts with typical recurrence periods of many
years (Tanaka $\&$ Shibazaki 1996). Most of the XN are believed 
to harbor a Black Hole (BH)
as the compact object. Because of large changes in the 
effective accretion rate
during the outburst, XN often pass through some of the spectral states
that have been identified in BH binaries (e.g., McClintock \& Remillard 2006).  
This gives us the possibility to study the physical 
properties of emitting
regions (disk, corona, jets) and their evolutions. 
The two main spectral states are the
Thermal Dominant State (TDS) and the Low/Hard State (LHS). They are
characterized by different combinations of the soft (multicolor black-body
coming from an optically thick and geometrically thin accretion disk), hard
(power law, with a possible break at 50--100~keV) and reflection (continuum
excess, fluorescence lines and K-edge) spectral components. These 
characteristics are coupled to
different properties of variability, to Quasi Periodic Oscillations (QPOs)
observed in the power spectrum (e.g., Belloni 2001, 2005) and to radio
changes. Indeed, a compact jet is usually observed in the LHS while it is
quenched in the TDS (Corbel \etal 2000, 2003; Gallo \etal 2003, 2006). 
The jet can
contribute to the high-energy emission and models where the base of a compact
jet plays a major role have been proposed (Markoff \etal 2005).\\

At the onset of an XN outburst, the hard X-rays usually reach the maximum
first; after that, a soft component develops. Then, BH XN usually display the
TDS characterized by a power law (with $\Gamma$~$\sim$~2.5) extending to high
energies along with the strong ultra-soft component. Some sources tend to
evolve into the LHS in the late stages of the outburst (e.g., Cadolle Bel
\etal 2004). The spectrum is then dominated by a hard power law
($\Gamma$~$\sim$~1.5--2.0) with a break, attributed to 
the Comptonization of the low energy 
radiation by a hot electron cloud (or
corona/jet) with thermal distribution. The soft component, 
probably emitted by the disk, is very weak. In the past,
distinct states have been studied by combining data 
from several high-energy
instruments (e.g. Grove et al. 1998; Zdziarski et al. 2001; Rodriguez \etal
2003; Cadolle Bel \etal 2006). Larger samples 
of simultaneous observations 
over an extended energy range are necessary: they may 
help us to better understand the relation
between the different components, their links with the 
changes in the accretion
rate and the nature of any additional parameter 
that may drive the state changes (e.g., Homan et al. 2001; 
Rodriguez et al. 2003). In the
recent years, many BH Candidates (BHC) 
have undergone outbursts with no transitions, e.g.,
GS~1354$-$64 (Brocksopp \etal 2001) or XTE~J1118$+$480 (Remillard \etal 2005). 
Even some neutron stars did so (like Aquila X$-$1 in Rodriguez 
\etal 2006a): a systematic study of those systems 
may reveal important clues to understand 
the physical mechanisms of an XN outburst.\\

\sw~was discovered in hard X-rays with the {\it Swift}/Burst Alert Telescope
(BAT) on 2005 May 30 (Palmer et al. 2005). The {\it Swift}/X-ray Telescope
(XRT) observation revealed a variable source with a spectrum well fitted by a
power law of photon index $\Gamma$~$\sim$~2.1 and with a column density
$N_{\rm H}$~$\sim$~2.0~$\times$~10$^{21}$~atoms~
cm$^{-2}$ (Morris et al. 2005). 
This latter value is compatible with a location of \sw~
at several kpc 
without requiring intrinsic absorption. The source was also clearly detected
in UV with the {\it Swift}/UV Optical Telescope UVOT 
(Still \etal 2005). On the ground, the
optical counterpart found with the MDM 2.4~m telescope revealed a new star
within the {\it Swift} error circle (Halpern et al. 2005). At the beginning of
July, Fender \etal (2005) conducted a 8~hour radio observing run at a
frequency of 1.7~GHz (18~cm wavelength) with the Multi-Element Radio-Linked
Interferometer Network (MERLIN). They reported a probable point-like radio
counterpart which is consistent with a compact jet. 
In X-rays, the 1.2--12~keV source
flux increased to the maximum value of $\sim$~200 mCrab in few days (see
Figure~\ref{LCsimult} at MJD~$\sim$~53560) and then it 
started to decay slowly. The hard
power law spectrum observed with the {\it Swift}/XRT (Morris \etal 2005) and
the 0.6~Hz QPO detected in pointed {\it RXTE} observations are characteristic
of the LHS (Morgan \etal 2005; Ramadevi \& Seetha 2006). 
Preliminary {\it INTEGRAL} observations lead to the same 
conclusion (Cadolle Bel et al. 2005). Seven months after the observations shown in
this work (i.e., end of 2006 March), the same conclusion was 
reached by Miller \etal (2006) who observed
the source up to 100~keV when it was fainter.\\

According to the characteristics mentioned above, 
\sw~may be a good BHC and it was therefore 
an excellent target for the {\it INTEGRAL} mission.
Consequently, we triggered our Target of Opportunity (ToO) campaign for XN 
in the Galactic Halo on August 10--12. This program targets XNs located
in regions out of the galactic disk and bulge where they are particularly
prone to be studied at different wavelengths because less subject to
absorption and source confusion. Preliminary results reported in Cadolle Bel
\etal (2005) showed that the source was then still in a LHS in the decay phase of the eruption. 
Thanks to the
large multi-wavelength program associated to the {\it INTEGRAL} 
ToO observation, we could also
trigger simultaneously optical (EMMI on NTT and ROSS on REM), 
Near Infra-Red (hereafter NIR) with REMIR (on REM) 
and radio (VLA) observations on August 11 for 3.6, 1, 7.3 
and 16.2 ks exposure times respectively.
Moreover, {\it RXTE} pointed observations ($\sim$3~ks) 
taken during our ToOs were available
so we added them in our analysis. We report here the 
results of our {\it INTEGRAL}
ToO observations on \sw~together with {\it RXTE} and 
optical/NIR/radio data,
starting with the description of the available data and of the analysis
procedures employed (Section~2). Our results are presented 
in Section~3 before their discussion and interpretations (Section~4).\\

\section{Observations and data reduction}

Table~\ref{log} reports the log of the performed observations of \sw~giving
for each of them the instrument data available, dates, exposure (per
instrument) and observing modes. The {\it INTEGRAL} observations were
performed on 2005 August 10--12 with a HEX dither pattern \citep{jen03}
consisting of series of seven pointings, one on the source and six around it
at 2$^\circ$ distance (along an hexagonal pattern). 
The period of our {\it
INTEGRAL} observations is also indicated in 
Figure~\ref{LCsimult}: it shows 
the 1.2--12~keV {\it RXTE}/ASM daily average light curve since the discovery
of the XN up to 2006 June 15. \\

\subsection{{\it INTEGRAL} data analysis}

We reduced the IBIS and JEM-X data with the standard analysis procedures of
the Off-Line Scientific Analysis {\tt OSA~5.1} released by ISDC, with
algorithms described in Goldwurm \etal (2003) and Westergaard \etal (2003) for
IBIS and JEM-X respectively. Only the JEM-X1 monitor operated during our
observation. Following recommendations of the {\tt OSA~5.1} release,
IBIS/ISGRI events were selected to have corrected energies $>$~15~keV and rise
time channels between 7 and 80. For the background corrections, we used the
set of IBIS/ISGRI (default) maps derived in 256 energy channels from empty
field observations. For the off-axis correction, maps and response matrices, 
we used those of the official {\tt OSA~5.1}~release; all the maps were rebinned
to our chosen energy bins. We took the latest available response files of {\tt
OSA~5.1} which are time dependent and based on recent Crab observations. In
the analysis, we considered the presence of other sources 
which were active in the considered region. 
Apart from the {\it Swift} XN no source was detected above
80~keV; only the neutron star binary system 
4U~1812$-$12 was active above 40~keV. We
extracted light curves and spectra in different energy bands 
between 17--20 and
500~keV and we built light curves and Hardness Ratios (HR). The SPI data were
preprocessed with {\tt OSA~5.1} using the standard energy calibration gain
coefficients per orbit. The {\tt spiros~9.2}~release \citep{skin03} was used
to extract the spectra of \sw~with a background model proportional to the
saturating event count rates in the Ge detectors. Concerning the instrumental
response, versions 17 of the IRF (Image Response Files) and 4 of the RMF
(Redistribution Matrix Files) were taken as our ToO took place after the
failure of the two detectors.\\

\subsection{{\it Rossi-XTE} data analysis}

We reduced the {\it RXTE} data using the {\tt LHEASOFT v6.0.4} package,
following the standard methods described in the Cook Book in a way similar to
what we have done in Rodriguez \etal (2003, 2006b). 
We defined the ``basic'' Good
Time Intervals (GTI) as the times during which the offset pointing was less
than 0.02$^\circ$ from the source and the elevation above the Earth limb
higher than 10$^\circ$. We also required that GTI correspond to times when
both the Proportional Counter Units (PCU) 
0 and 2 were turned on. This resulted in 3104~s of good times.\\

We extracted Proportional Counter Array (PCA) spectra and 16~s light curves
from standard 2 data of the top layers of PCU 0 and 2. We generated background
spectra with {\tt pcabackest~v3.0} using the calibration files available for
bright sources. Response files were generated with {\tt pcarsp~v10.1}. Due to
uncertainties in the PCA response, 0.6$\%$ systematics were added to all
channels. In order to study the rapid temporal variability of \sw, we also
extracted high resolution light curves from Event data with a time 
resolution around 500~$\mu$s. 
This allowed us to study the timing properties up to
$\sim1000$~Hz. We restricted the light curve extraction to absolute channels
5--121 (2--52 keV) in order to limit effects of the 
background at low and high energies.\\

Finally, we extracted spectra from the High Energy Timing Experiment (HEXTE)
in an identical way of that presented in Rodriguez \etal (2003): for each
cluster (A and B), the ``on'' from the ``off'' source pointings were separated
before extraction of source plus background and background spectra, that were
further corrected for deadtime. Spectral responses were generated with {\tt
hxtrsp~v3.1}.

\subsection{Optical and NIR data analysis}

Our ToO observations were carried out as part of the European Southern 
Observatory (ESO) program ID $\#$075.D-0634 (PI S. Chaty). 
On 2005 August 11 (around U.T. 02) 
we obtained optical photometry
in $B$, $V$ $R$ and $I$ bands with the spectro-imager EMMI installed on the
New Technology Telescope (NTT) at La Silla Observatory. 
We used the large field imaging of EMMI's
detector which gives an image scale of 0\farcs166 pixel$^{-1}$ and a field of
view of 9\farcm9$\times$9\farcm0. Concerning the photometric observations, we
took an integration time of 60~s for each exposure. We observed four
photometric standard stars of the optical standard star catalog of Landolt
\etal (1992): T Phe A, B, C and D. Rapid photometric observations 
were performed in the $V$
band during the time interval between MJD~53593.17--53593.24. They 
were simultaneous to
the radio observations: 200 images of 1~s exposure time were obtained every
27~s, covering a total of 93 minutes. We detected a scattering of 0.1
mag (within the error bars) in the course of this rapid photometry. 
Further observations were performed with the REM 
telescope (Zerbi et
al. 2001; Chincarini et al. 2003; Covino et al. 2004) equipped with the
ROSS optical spectrograph/imager and the REMIR NIR camera.
Observations were carried out on August 11.98, 17.99 and 27.04 (U.T.) in
the optical ($R$ filter) and on August 11.04 (U.T.) in the NIR 
(standard $J$, $H$ and $K_s$ filters). NIR photometric observations were
calibrated against the $2MASS$ catalog.\\

We used the {\tt IRAF} (Image Reduction and Analysis Facility) suite to
perform data reduction, carrying out standard procedures 
of optical and NIR image
reduction, including flat-fielding and subtraction of the 
blank NIR sky. We carried out aperture photometry and
we then transformed instrumental magnitudes into apparent magnitudes with the
standard relation: $mag_{\rm app} = mag_{\rm inst} - Z_{\rm p} - ext * AM$
where $mag_{\rm app}$ and $mag_{\rm inst}$ are respectively the apparent and
instrumental magnitudes, $Z_{\rm p}$ the zero-point, $ext$ the extinction and
$AM$ the airmass. The observations were performed through an $AM$ close to
1.\\

We also carried out optical spectroscopy with EMMI. We took 12 spectra with
the grisms $\#1$ covering 4000 to 10000~\AA~with a resolution between 200 and
500~\AA~(depending on the wavelength). Each spectrum had an exposure time of
300~s, giving a total integration time of 60~min. In order to extract spectra
and perform wavelength and flux calibrations, we used the {\tt IRAF
noao.twodspec} package.\\

\subsection{Radio data analysis}

We observed \sw~with the NRAO\footnote{The National Radio Astronomy
Observatory is a facility of the National Science Foundation operated under
cooperative agreement by Associated Universities, Inc.} Very Large Array (VLA)
at 1.4, 4.9, 8.5 and 15~GHz (respectively 21, 6, 3.5 and 2.0~cm wavelength) on
2005 August 11 from 4:20 to 8:50~U.T. (average MJD~53593.28) 
with the VLA in its
C configuration. The receiver setup included two IF pairs of 50~MHz bandwidth
each. We obtained 20~min snapshots at 1.4, 4.9 and 15~GHz. We also derive a
$\sim$2.5 hour lightcurve at 8.5~GHz. The observations were conducted as
follows: 10~min scans on \sw~preceded and followed by 1~min scans on the VLA
phase calibrator \object{J1743$-$038} (located 3.4\degr\ apart). The primary
flux density calibrator used was \object{J0137+331} (\object{3C~48}). The data
were reduced using standard procedures within the NRAO {\tt aips} software
package.

\section{Results of the analysis}

\subsection{X-ray position}

\begin{figure}[t!]
\centering\includegraphics[width=0.7\linewidth]{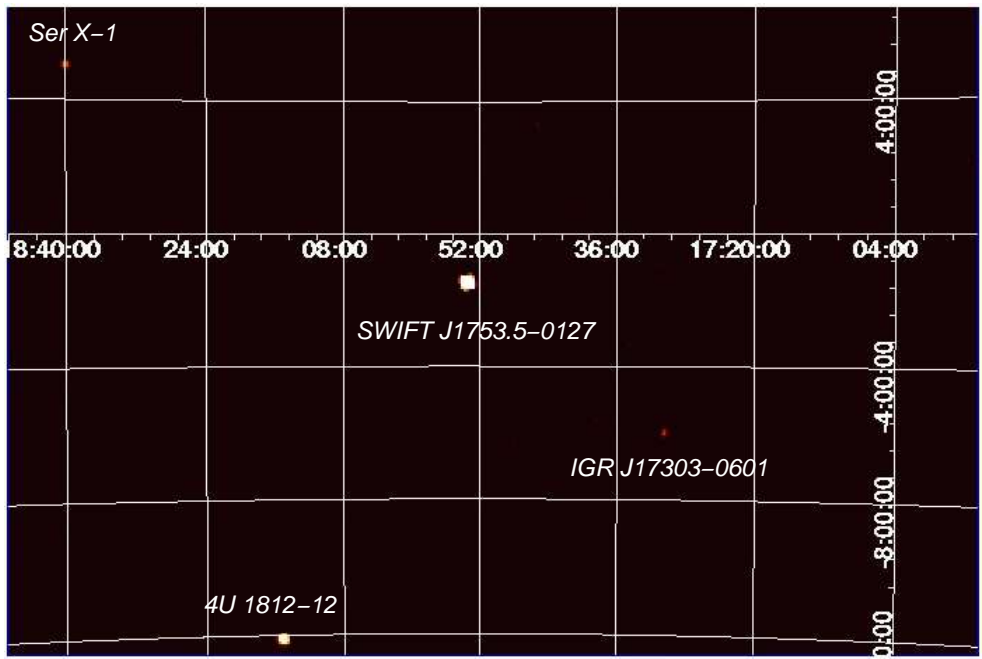}
\caption{\label{Mosa}The IBIS/ISGRI reconstructed sky image of the
region around \sw~in the 20--40 keV band. \sw~appears at a
significance level of 440$\sigma$ over the background.
The other sources in the image are the neutron star 4U~1812$-$12
(at 34$\sigma$), the X-ray burster Ser~X$-$1 (14$\sigma$) and 
the {\it INTEGRAL} obscured source IGR~J17303$-$0601 (12$\sigma$).}
\end{figure}

In the combined IBIS/ISGRI images obtained during the 
hard outburst, the transient source \sw~was
detected at 440, 320, 174 and 22$\sigma$ respectively in the 20--40, 40--80,
80--160 and 160--320~keV energy 
bands. Figure~\ref{Mosa} shows the 20--40~keV IBIS/ISGRI mosaic:
three other sources were detected, the neutron 
star 4U~1812$-$12 (34$\sigma$), 
the X-ray burster Ser~X$-$1 (14$\sigma$) and 
the {\it INTEGRAL} obscured source named 
IGR~J17303$-$0601 (12$\sigma$). The best-fit
position found by IBIS/ISGRI over $\sim$ 176~ks exposure in the 20--40~keV
mosaic sky image was $\alpha_{\rm J2000}=17^{\rm h}~53^{\rm m}~28\fs4$ 
and $\delta_{\rm J2000}=-01\degr~27\arcmin~17^{\prime\prime}$, with an
accuracy of $14^{\prime\prime}$ at the 
$90\%$ confidence level (Gros \etal 2003). 
This position is compatible with the {\it Swift}/XRT location (Burrows \etal
2005) as the angular separation between ISGRI and XRT 
is $7^{\prime\prime}$ only. It is also 
consistent with the most precise position of \sw~derived from the {\it
Swift}/UVOT data (Still \etal 2005) since the offset is only
$11^{\prime\prime}$ (within the error circle). The high-energy 
source and the UV/optical/radio counterparts are 
therefore unambiguously associated to
the X-ray transient source.\\

\subsection{X and $\gamma$-ray light curves}

\begin{figure}[t!]
\centering\includegraphics[width=1\linewidth]{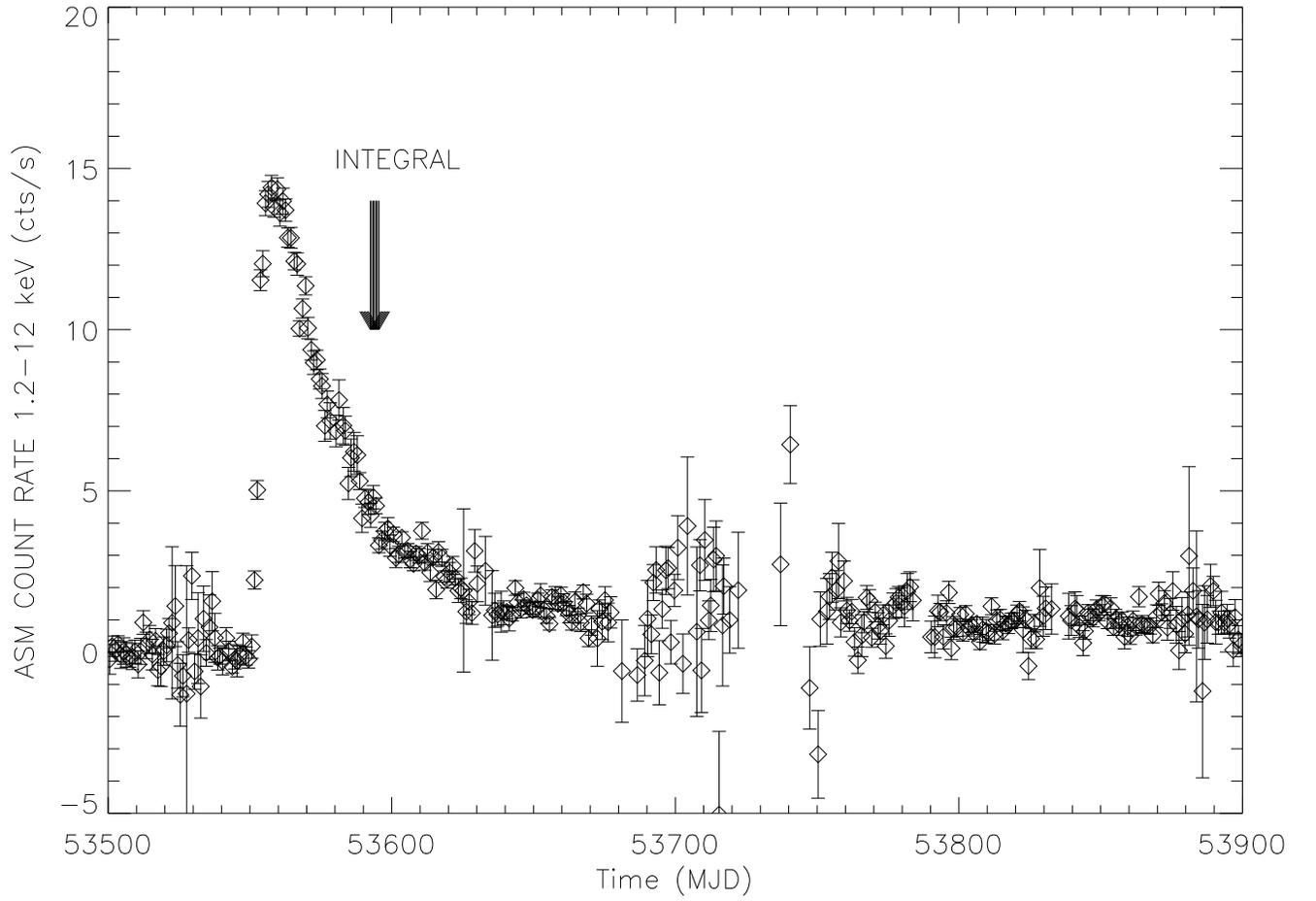}
\caption{\label{LCsimult}{\it RXTE}/ASM daily average
(1.2--12~keV) light curve of \sw~from 2005 mid-May up
to 2006 mid-June (MJD~=~JD~$-$~$2 400 000.5$)
with the periods of our simultaneous {\it INTEGRAL}
observations indicated by an arrow (its 
width illustrates the {\it INTEGRAL} range). The gap corresponds 
to the passage of the Sun close to the source. 
Error bars are at 90$\%$ confidence level.}
\end{figure}

Figure~\ref{LCsimult} shows the 1.2--12 keV {\it RXTE}/ASM daily 
average XN light
curve from just before its discovery, during its outburst rise and peak
(in 2005 July) and down to its return to a lower ASM level 
(almost undetectable
around 2006 mid-June). 
From the end of 2005 May up to July 9, the ASM average count
rate (where 1~Crab~=~75~cts~s$^{-1}$) increased and its flux reached the
maximum value of $\sim$ 200~mCrab (MJD~53560). The flux 
then decreased to $\sim$
14~mCrab (2005 October 10, MJD~53650) before 
the source became nearly undetectable by the ASM.
Assuming an exponential shape for both the rise and the decay phases seen in 
the ASM, we obtained time constants of 
4.5~$\pm$~0.2 (rise) and 32.0~$\pm$~0.2 (decay)
days respectively. The characteristic decay time we derived is compatible with
the usual behavior of XN 
in outburst (Tanaka \& Shibazaki, 1996; Chen et al.
1997) like, e.g., XTE~J1720$-$318 (Cadolle Bel \etal 2004).\\

\begin{figure}[t!]
\centering\includegraphics[width=1\linewidth]{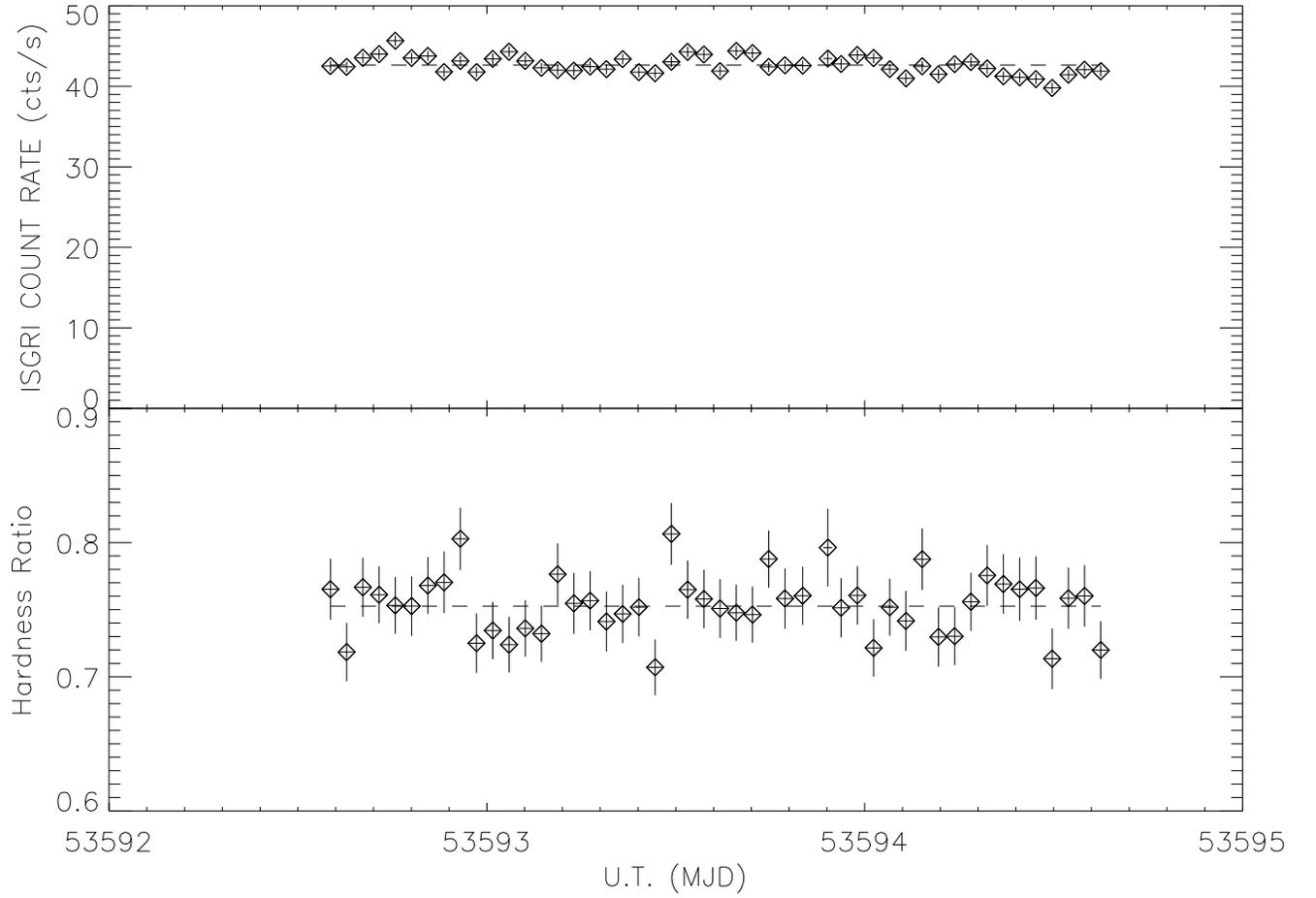}
\caption{\label{LCHR}The 20--320 keV IBIS/ISGRI light curve of \sw~and
corresponding HR between the 40--80 and the 20--40 keV significant energy bands
(average levels denoted by dashed line) during our ToO
{\it INTEGRAL} observations. Error bars are at 90$\%$ confidence level.}
\end{figure}

The large arrow overplotted on Figure~\ref{LCsimult} indicates 
the interval covered by the dedicated
{\it INTEGRAL} ToO observations discussed in this work, 
for which we also triggered
simultaneous optical, NIR and radio observations. 
Between MJD 53592--53594.4,
the IBIS/ISGRI count rate was almost constant at 43~cts~s$^{-1}$
(Figure~\ref{LCHR}, top) which corresponds to a significative flux of $\sim$
205~mCrab between 20--320~keV. Figure~\ref{LCHR} (bottom) also shows that the
HR between the 20--40 and 40--80 keV energy bands was compatible with being
constant ($\sim$0.75). The source was observed in the decay phase of its 
outburst which started (as seen in Figure~\ref{LCsimult}) around MJD
53560. However, the source was still bright in the {\it INTEGRAL}/IBIS energy
band at the time of our observations. 
This favors the scenario that the source probably stayed in the LHS
during (see also Ramadevi \& Seetha 2006) and after its outburst 
(see Miller \etal 2006). In
particular, \sw~did not show transitions to a softer state: 
we will discuss this issue in Section 3.4.

\subsection{Timing variability from high-energy data}

\begin{figure}[t!]
\centering\includegraphics[width=1\linewidth]{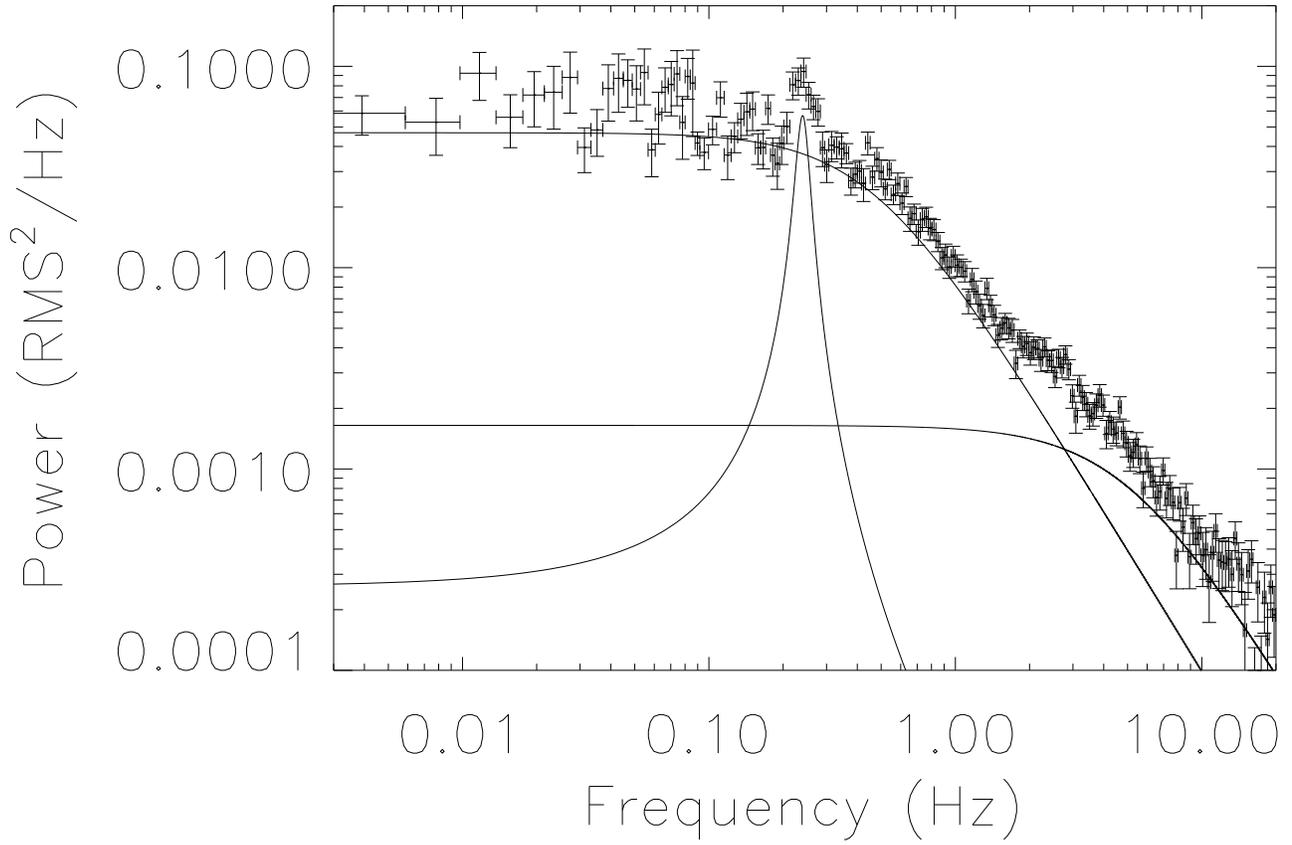}
\caption{\label{qpo} Power Density Spectrum of 
\sw~obtained with the {\it RXTE}/PCA data
on 2005 August 11. A QPO frequency of 0.24~Hz is found; an F-test 
confirms its necessity at the 99.95$\%$ level. The lines
correspond to the three Lorentzians which are necessary to best fit the data
(see Table~\ref{PDS}): we obtain a \kir~of 1.20 (163 dof).}
\end{figure}

We produced a Power Density Spectrum (PDS) from the PCA high time resolution
light curves with {\tt POWSPEC~V1.0}. The PDS was generated on several
intervals of 256~s between 3.9$\times10^{-3}$ and 1024~Hz. The results were
averaged together and the resulting PDS is shown in Figure~\ref{qpo}. It was
corrected for white noise before fitting. Since above 20~Hz the PDS is
compatible with Poisson noise, we restricted our fitting to the
$3.9\times10^{-3}$~--~20~Hz frequency range. The continuum is well represented
by the sum of 2 broad zero centered Lorentzians while an additional third one
is needed to account for a QPO around 0.24~Hz. Even if it is faint (5.4$\%$
r.m.s.), an F-test confirms that this component is required (99.95$\%$ level
probability). Note that in the case of a broad Lorentzian, the frequency at
which the feature attains its maximum (in a $\nu P_{\rm \nu}$ representation)
is defined by $\nu_{\rm max}~=~\sqrt{{\nu_{\rm 0}}^2+\Delta^2}$ (Belloni \etal
2002) where $\Delta$ is the half width at half maximum. Hence in our case this
leads to $\nu_{\rm max}~=~\Delta$. This model of 3 Lorentzians provides a
relatively good description of the PDS with a $\chi^2~=~197.1$ for 163 degrees
of freedom (dof). The normalized PDS (after Leahy \etal 1983) and the
individual components are represented in Figure~\ref{qpo}. The parameters of
each Lorentzian are reported in Table~\ref{PDS}. The QPO mean value is lower
than the 0.6~Hz QPO reported after the peak of the outburst 
(Morgan \etal 2005; Ramadevi \& Seetha 2006). 
This trend (decrease of the QPO frequency during the decrease of the
eruption) is sometimes observed in other BHCs and has been associated to the 
increase of the inner radius of the accretion disk 
(Kalemci \etal 2001, 2002; Rodriguez \etal
2002, 2004a; Belloni \etal 2005). However, we can not exclude another
interpretation for the QPO based on the pulsation modes of the corona
(Shaposhnikov \& Titarchuk 2006). We did not detect the QPO in either JEM-X
nor ISGRI data: this is probably due to their lower collecting area than PCA
and to the faintness of the QPO at the time of our {\it INTEGRAL} ToO. In
any case the high level of band-limited noise ($\sim$ 27$\%$ r.m.s., see
Table~\ref{PDS}) we observed is typical of the LHS of XN.\\

\subsection{X and $\gamma$-ray spectral results}

Figure~\ref{LCHR} shows the flux and HR measured during the {\it INTEGRAL}
high-energy observations. The IBIS/ISGRI flux appears to be constant as well
as the HR ($\sim$0.75) during the course of our ToO observations. As there is
no significant variation in the HR, we therefore used the whole data from
JEM-X, IBIS/ISGRI and SPI of this hard outburst to build up an average 
spectrum on a wide band together
with the simultaneous PCA and HEXTE data obtained during our {\it INTEGRAL}
ToO. We added 3$\%$ systematic errors for JEM-X (in the 5--25~keV range) 
and SPI (in the 22--600~keV range) and 2$\%$ for IBIS
(in the 18--320~keV range). Using {\tt XSPEC v11.3.0}
(Arnaud \etal 96), we fitted the resultant spectra simultaneously with PCA
(3--25~keV) and HEXTE (20--100~keV). In order to account for uncertainties in
the cross-calibration of each instrument, a multiplicative constant was added
in the spectral fits to each instrument data set: it was 
frozen to 1 for PCA
and set free for JEM-X, IBIS/ISGRI, SPI and HEXTE.\\

Following the approach described in Cadolle Bel \etal (2006) when modelling
the LHS spectra of Cygnus~X-1, we fitted our simultaneous
X/$\gamma$-ray data from the {\it RXTE} and {\it INTEGRAL} 
satellites to different spectral distribution functions. 
We added progressively each spectral component 
to account for possible 
presences of Comptonization (Titarchuk 1994), 
multicolor black body disk (Mitsuda \etal 1984), reflection or Fe line. 
We then tested if each component was required by the 
improvement in \kir. We 
always used a fixed absorption column density~$N_{\rm H}$ of
2~$\times$~10$^{21}$~atoms~cm$^{-2}$, 
taking the value obtained with our optical
data (see Section 3.5). Indeed, we can not constrain 
it with the PCA nor the JEM-X data. First, we
tried to fit simultaneously all the data with the Comptonization
model of Titarchuk (1994) convolved by absorption ( {\sc
cons}*{\sc wabs}*{\sc comptt} in {\tt XSPEC} notation). We obtained a seed
photon temperature $kT_{\rm 0}$ of $0.51~\pm~0.08$~keV, a plasma temperature
$kT_{\rm e}$ of $88 \pm 14$~keV and a plasma optical depth $\tau$ of $0.67 \pm
0.14$. However, this model gave 
an unacceptable fit (\kir~=~2.14 with 322 dof) to the data.
Including a multicolor black body disk did not improve the fit,
with most of the large residuals showing up below 20 keV between 
the PCA and JEM-X data sets and in the HEXTE data 
between 20--30~keV and above 70~keV. The reflection 
\citep{mag95} is usually employed to account for a soft excess around
10~keV. This latter component (noted {\sc
reflect} in {\tt XSPEC}) models the X-ray reflection of the comptonized
radiation from neutral or partially ionized matter, presumably the optically
thick accretion disk (see also Done \etal 1992; Gierli\'nski 
\etal 1997, 1999). Adding this component with an inclination
angle equal to 63$^\circ$ (default), we obtained a significant decrease in
\kir~(1.60) but a very high plasma temperature. This model gave an
improved fit but, again, it was still unacceptable.\\

\begin{figure}[t!]
\centering\includegraphics[width=0.6\linewidth,angle=270]{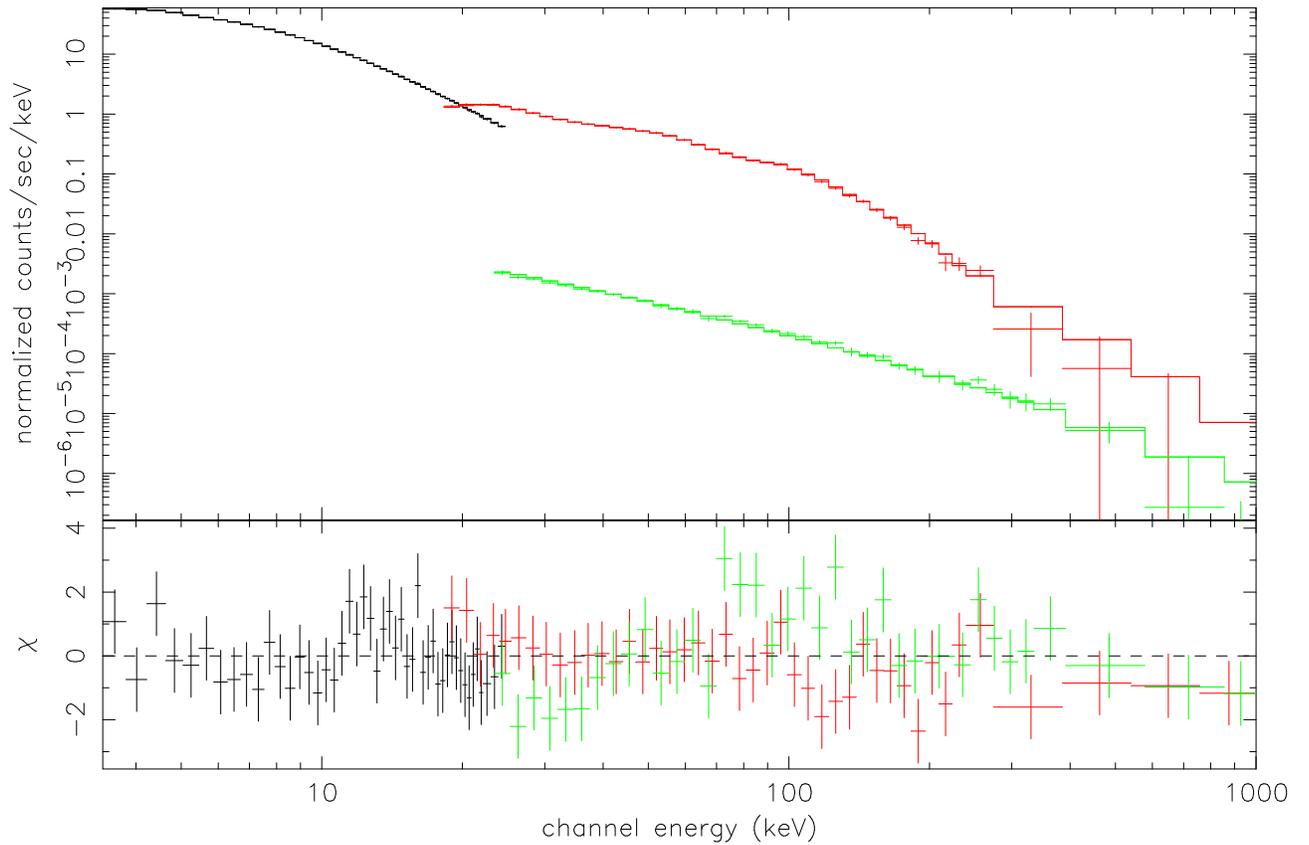}
\caption{\label{tot} Spectra of \sw~during our {\it INTEGRAL}
ToO with the PCA (black),
IBIS (red) and SPI (green) data along with
the best-fit model: absorbed thermal Comptonization convolved 
by reflection (see text). Residuals in $\sigma$ units are also 
shown.}
\end{figure}

\begin{figure}[t!]
\centering\includegraphics[width=0.8\linewidth]{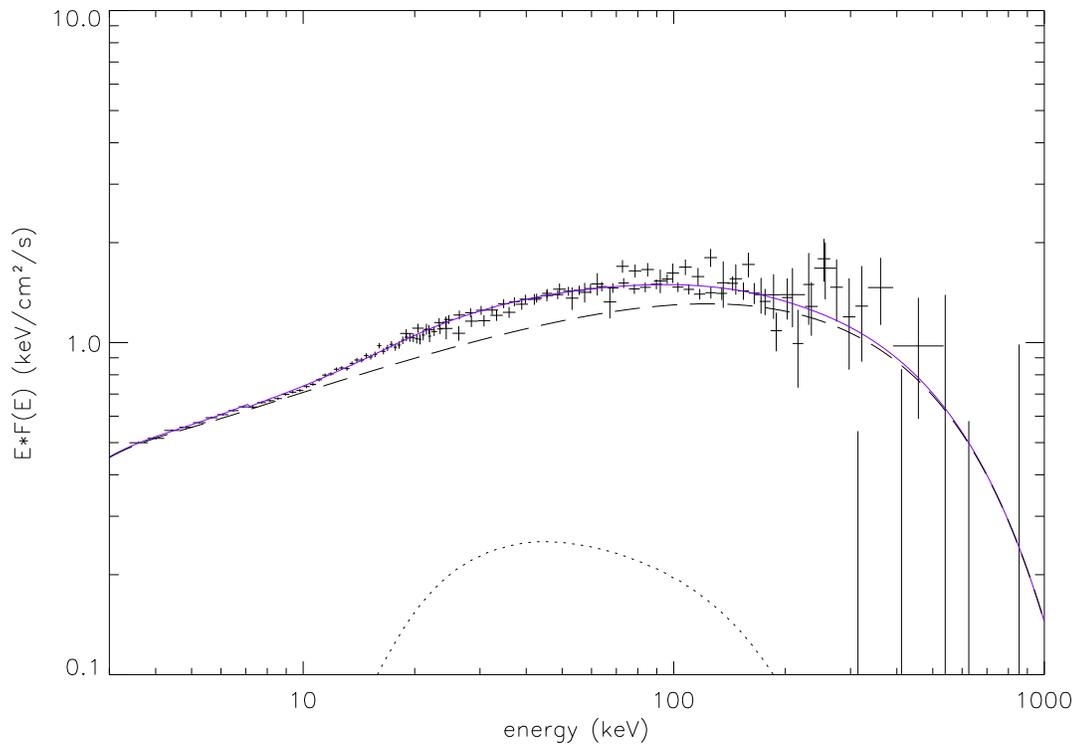}
\caption{\label{totefe} $EF(E)$ spectra (units keV~cm$^{-2}$~s$^{-1}$) 
of \sw~during our {\it INTEGRAL}
ToO with the PCA, IBIS and SPI data along with
the best-fit model (thick): absorbed 
Comptonization (dashed) convolved by reflection (dotted).}
\end{figure}

As the JEM-X and HEXTE instruments overlap the energy 
bands of, respectively, PCA and IBIS/ISGRI (and SPI) and 
as these latter instruments are supposed to be the best 
calibrated, we tried to fit simultaneously only
the data from PCA, IBIS/ISGRI and SPI. 
With the {\sc cons}*{\sc wabs}*({\sc reflect}*{\sc comptt}) 
model, we obtained a reasonable \kir~of 1.17
(with 121 dof). Our best-fit model has the following 
parameters: a seed photon temperature
$kT_{\rm 0}$ of 0.54$_{-0.07}^{+0.04}$~keV, an electron temperature $kT_{\rm
e}$ of 150~$\pm$~26~keV, an optical depth $\tau$ of 1.06~$\pm$~0.02 and a
reflection fraction 
$\Omega/2\pi$~=~0.32~$\pm$~0.03. Normalization constants between
IBIS/ISGRI and SPI were respectively equal to 0.97 and 1.20 when PCA was frozen
to 1. Note that these parameters (even if the plasma temperature is higher
when the reflection component is added) are compatible with the ones obtained
previously while fitting simultaneously all the data 
(from PCA, JEM-X, HEXTE, IBIS/ISGRI and SPI). This 
best-fit model over-plotted on the data
from PCA, IBIS/ISGRI and SPI is reported in count units in Figure~\ref{tot}. 
We have rebinned the IBIS/ISGRI (above $\sim$300~keV) and SPI (above
$\sim$500~keV) data at the level of 3$\sigma$. Figure~\ref{totefe}  
also shows our
best-fit model with its components 
overplotted on the same data in $EF(E)$
units (keV~cm$^{-2}$~s$^{-1}$).\\

The parameters we derive are compatible with the source being in the LHS. 
The relatively high plasma temperature indicates that the Comptonizing medium 
(corona) remains hot: this is probably due to the weak disk emission unable 
to efficiently cool down the corona by Compton scattering. 
The derived value ($\sim$0.33) of the Comptonization parameter $y$, 
given by $kT_{\rm e}/m_{\rm e}c^2$~Max ($\tau$, $\tau^2$), 
is typical of a LHS, as fully discussed in Cadolle Bel \etal 
(2006) for Cygnus~X-1. 
We therefore conclude that the source was very hard 
during the eruptive phase with a
behavior and spectral Comptonization parameters typical of a BH in LHS.

\subsection{Optical and NIR results}

Concerning the photometry obtained on 2005 August 11 (U.T. 02) with EMMI, the
apparent magnitudes we derived are: $B=16.73 \pm 0.02$, $V=16.46 \pm 0.02$,
$R=16.15 \pm 0.02$ and $I=15.64 \pm 0.03$. They are nearly 0.5 mag fainter
than the apparent magnitudes obtained by Still \etal (2005) and Torres \etal
(2005a, b) during the period of July 1--11, probably because our observations
were obtained later in the outburst. Torres and co-workers suggested a slow
change rate at a level of 0.1 mag per week. 
This is exactly the behavior we have
observed later and it is compatible with the evolution of an XN in outburst. 
The same trend was observed during subsequent 
monitoring with REM: indeed, the flux in the $R$ filter results to be 
almost constant at the average level of $\sim$ 16.45 
mag on August 17 and of $\sim$ 16.60 on August 27. REMIR 
photometry reveals a NIR counterpart with 
$J = 15.35~\pm~0.06$, $H = 15.11~\pm~0.07$ and 
$K = 14.73~\pm~0.15$ mag on August 11.\\

\begin{figure}[t!]
\centering\includegraphics[width=0.8\linewidth]{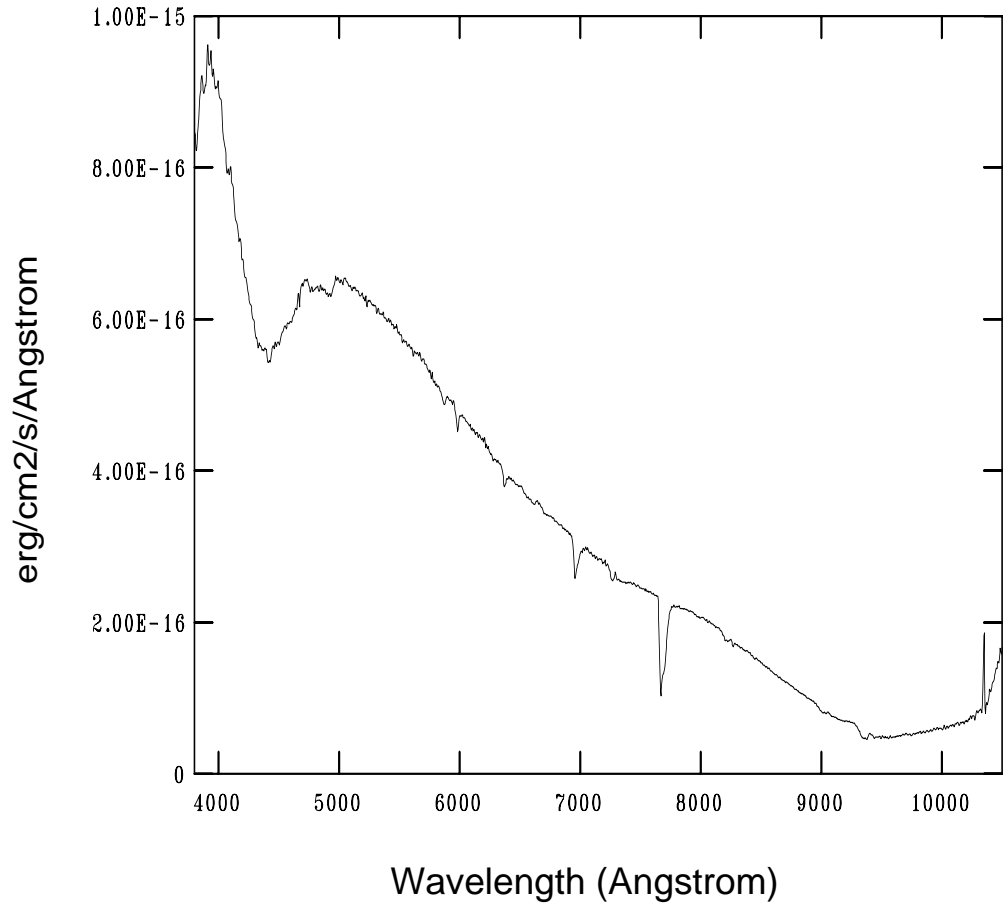}
\caption{\label{opt} Calibrated optical spectrum of \sw~(erg
cm$^{-2}$ s$^{-1}$ \AA$^{-1}$) obtained
on 2005 August 11 around U.T. 02 (see Table 1 for accurate dates) 
with the NTT telescope (see text for line details).}
\end{figure}

Figure~\ref{opt} represents the derived EMMI spectrum in
erg~cm$^{-2}$~s$^{-1}$~\AA$^{-1}$ units and calibrated with the EG274 standard
spectrophotometric star: the continuum is blue, suggesting the emission from
an accretion disk, and it shows many telluric lines at respectively 6300,
6900, 7200, 7700 and 8300 \AA. Interstellar lines at 5800, 5900 (NaI Doublet)
and 6300~\AA~are present. We do not observe any 
clear absorption nor emission lines 
apart from the H$\alpha$ (around 6500 \AA) and the 
\ion{He}{1} (around 5800, 6600 and
7200~\AA) lines: they are visible in 
absorption (but very faint). They probably 
emanate from the accretion disk as suggested 
by the blue continuum. 
The non-detection of any emission line differs from the results of 
observations reported earlier
during the outburst on July 3 (Torres \etal 2005a): the spectra showed then a
blue continuum with H$\alpha$ being the only emission line detected in the
spectrum with an Equivalent Width (EW) of $\sim~3~$\AA~and a Full Width at
Half Maximum (FWHM) of 2000~km~s$^{-1}$. The NaI doublet at 5890/96 \AA~was
also evident with EW~$\sim$~1.5~\AA. One week after (Torres \etal 2005b), a
broad double-peaked H$\alpha$ emission was detected each night (July 7, 10, 11
and 12) with mean EW of 3~\AA, mean double-peacked separation of
1200~km~s$^{-1}$ and a 
FWHM around 2000~km~s$^{-1}$. This is characteristic of
the outer accretion disk in X-ray binaries with a low mass companion star.
\ion{He}{2} (4686 \AA) emission was highly variable both in terms of strength
and profile shape, ranging between $\sim$~3 and $\sim$~5 \AA~in EW. On August
11, the H$\alpha$ line has almost disappeared and is not anymore
double-peaked: the evolution of this line is typical for LMXB (novae) in
outburst. The low amplitude and EW of the lines suggest that the disk has a
fainter contribution at the time of our simultaneous optical/{\it INTEGRAL}
observations, as also confirmed by the fact that the $B-V$ value stayed almost
constant (at 0.3 mag) during and after the outburst. Comparisons of the
spectra obtained between July and mid-August show the expected behavior of
LMXBs in eruption: bright contribution of a disk in optical, followed by a
decrease of this contribution simultaneously to a decrease of the soft X-ray
flux.\\

Finally, we could determine the column density along the line of sight, using
EW of the NaI Doublet (1.36~$\pm$~0.15~\AA). According to Munari \& Zwitter
(1997), $E_{\rm B-V}=0.25 \times$ EW$_{\rm NaI}$: therefore we obtain $E_{\rm
B-V}=0.34 \pm 0.04$~mag. Using the formula of Bohlin et al. (1978) 
$N(H_I+H_2)=5.8~\times~10^{21}~\times~E_{\rm B-V}$, we calculate $N_{\rm H}$
and we get $1.97~\pm~0.23~\times~10^{21}$~atoms~cm$^{-2}$. This value is well
consistent with the absorption determined by {\it Swift}/XRT (Morris \etal
2005). Depending on which total galactic absorption we get
($\sim$1.7~$\times~10^{21}$~atoms~cm$^{-2}$ from Dickey \& Lockman 1990
commonly used by the X-ray community or
$\sim$2.7$\times~10^{21}$~atoms~cm$^{-2}$ with Schlegel \etal 1998), the 
source $N_{\rm H}$ is below or comparable within errors with the total
galactic column density in this direction so the source can not be located at
more than 10~kpc. Besides, 
it is situated at least at more than 2--3~kpc in order to
be compatible with its absorption and its 
high latitude value ($l$=12.9$^\circ$). Therefore, a
probable distance for the source is between 4 and 8~kpc: we took 
the average value of 6~kpc in the following.

\subsection{Radio results}

We produced images using a natural weighting scheme. Considering the angular
resolution of the VLA (the synthesized beam), we detected at all observed
frequencies a point-like radio counterpart (angular radius $< 4\arcsec$) at a
position compatible with the MERLIN one. The obtained flux density at each
frequency ($S_\nu$) and the fitted spectral index $\alpha$ (where
$S_\nu\propto\nu^{+\alpha}$) in different frequency ranges are quoted in
Table~\ref{tabradiospec}. All errors are at the 1$\sigma$ level and we show
in parentheses the expected r.m.s. sensitivities for the corresponding amounts
of observing time. The discrepancy between the expected (0.04) and observed
(0.15) error values at 1.4~GHz is because we could only image the source
properly after excluding the galactic diffuse emission detected on short
baselines (hence increasing the noise). The discrepancy at 4.9~GHz is due to
the fact that we considered the higher error given by the {\tt jmfit}
procedure (Gaussian fitting). The source was slightly variable during our long
8.5~GHz observing run (see below), and different flux densities were obtained
depending on the measuring method: flux density within a box ({\tt imstat}),
peak and integral flux of a Gaussian fit ({\tt jmfit}). Therefore we give the
average value of 0.7$\pm$0.1. Finally, the VLA data at 15~GHz
could only provide a 3$\sigma$ upper limit of 0.60 mJy, which will not be
considered hereafter in our analysis and interpretation.\\

\begin{figure}[t!]
\center
\resizebox{1.0\hsize}{!}{\includegraphics{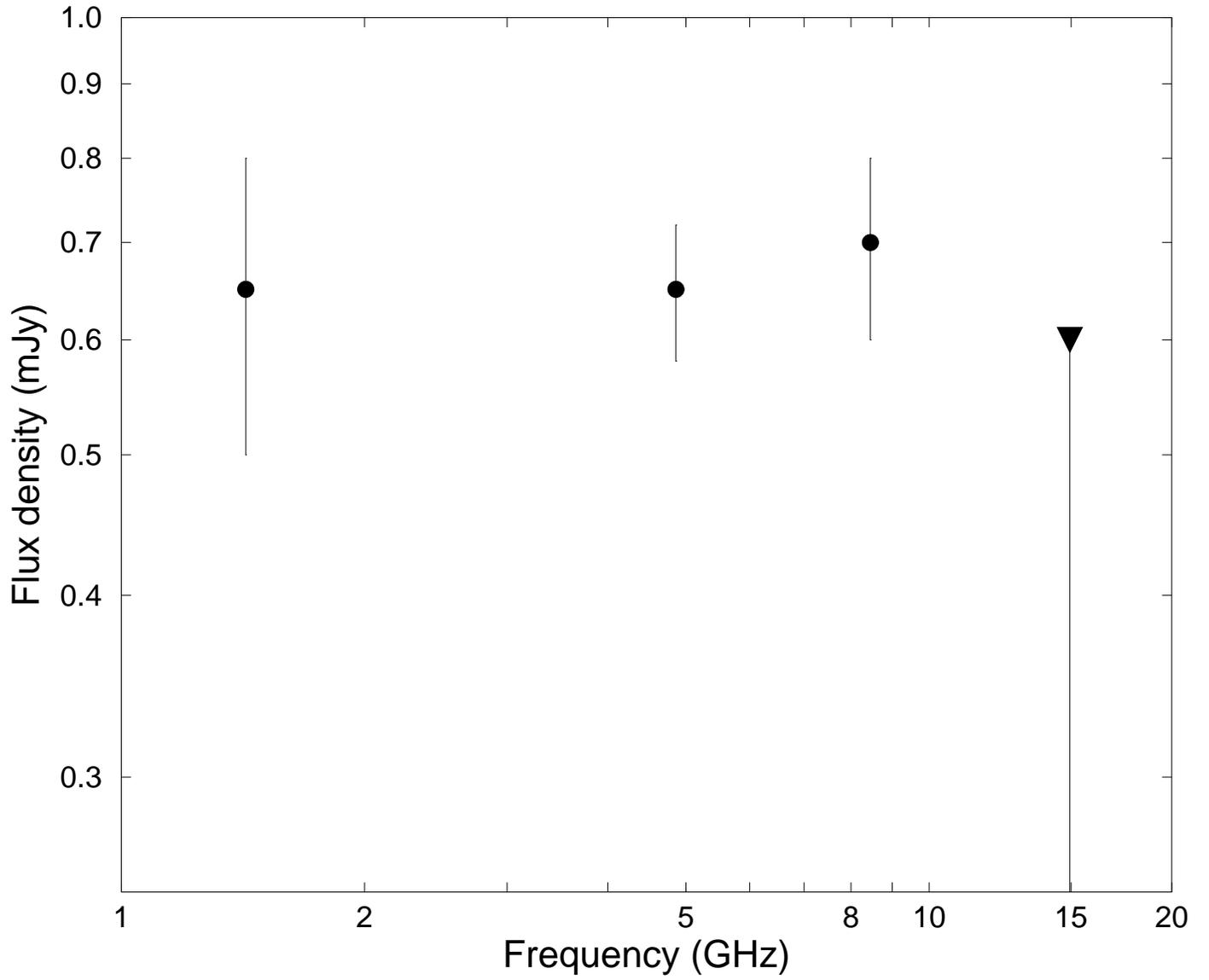}}
\caption{Radio spectrum of \sw~from 1.4 to 15~GHz. 
Error bars are at the 1$\sigma$ level while we show a 3$\sigma$ upper limit at 15~GHz. 
The data are compatible with a flat radio spectrum, interpreted as synchrotron
radiation from a partially self-absorbed conical jet.}
\label{radiospec}
\end{figure}

\begin{figure}[t!]
\center
\resizebox{1.0\hsize}{!}{\includegraphics{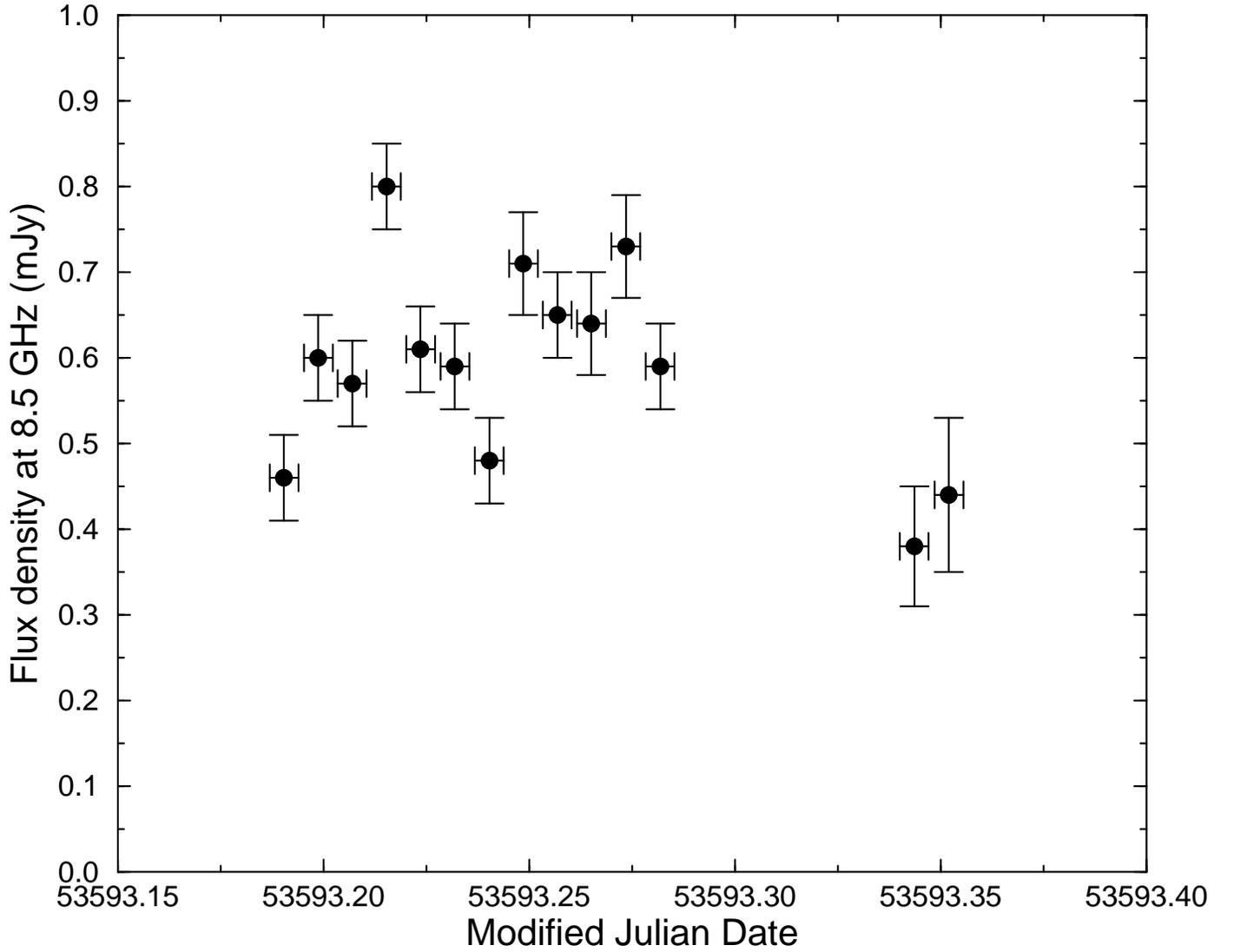}}
\caption{Radio lighcurve of \sw~obtained with the VLA at 8.5~GHz. 
The final two data points were acquired at low elevations and could have flux 
calibration errors higher than those shown. Even if excluding them, a $\chi^2$ 
test reveals that the data are not compatible with a constant radio flux density 
within the 99.5$\%$ confidence level.}
\label{radiolc}
\end{figure}

The flux densities as a function of observing frequency are shown in
logarithmic scale in Figure~\ref{radiospec}. A flat
-or slightly inverted- spectrum with $\alpha$=+0.03$\pm$0.03 is found after
excluding the 15~GHz upper limit point. 
We also show in Figure~\ref{radiolc} the radio
lightcurve obtained at 8.5~GHz in MJD units. The gap corresponds to the
observing time at the other frequencies. Each data point has been obtained
after measuring the flux density in an image produced with the corresponding
10~min snapshot. The final two data points were obtained after observations
conducted at low elevations (18 and 15\degr). Although the gain
corrections due to elevation have been taken into account, the flux density of
the phase calibrator also decreases sligthly 
in this last part of the observation, while the rms 
of its phases increases significantly.
Therefore, since the target flux density was 
too low to self-calibrate the phases, we consider 
that the measured flux densities and corresponding
uncertainties of these two data points could be both underestimated. The
average value of all the data points is $0.60 \pm 0.11$~mJy ($0.61 \pm
0.09$~mJy if we exclude the last two points). This value is lower but
compatible with the one obtained above after performing a single image with
all the data. This difference can be due to the sligthly variable behavior
seen in Figure~\ref{radiolc}. Indeed, even if excluding the last two data
points, a \ki test reveals that the data are not compatible with a constant
value at the 99.5$\%$ confidence level : this confirms that the radio flux  
was variable at the level of $\sim$30$\%$.

\section{Discussion}

\subsection{Spectral energy distribution}

\begin{figure}[t!]
\centering\includegraphics[width=0.6\linewidth]{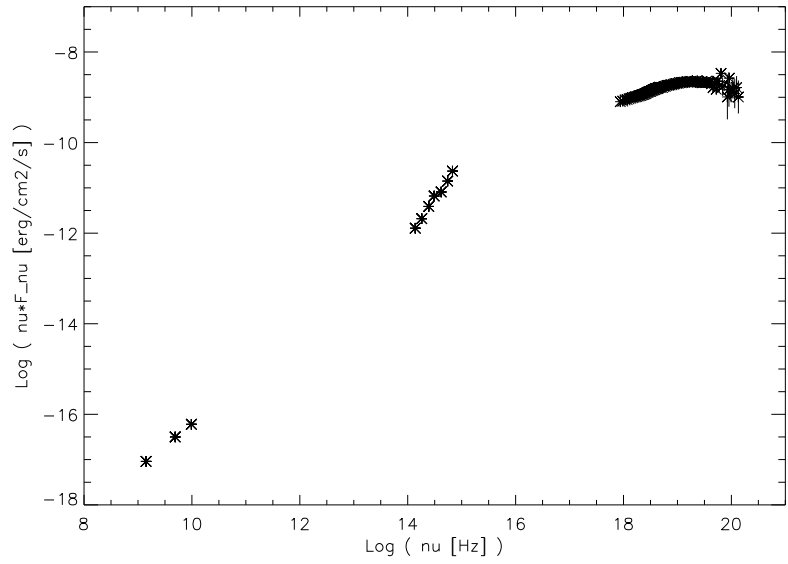}
\caption{\label{sed} Spectral energy distribution 
of \sw~on August 11 with radio (15~GHz upper limit data point excluded), IR,
optical and X-ray points (up to 535~keV) and errors. The NIR and
optical flux densities (filters $B$, $V$, $R$, $I$, $J$, $H$ and $K$) were
dereddened with $A_V$=~1.05$\pm$0.12 mag (see text). 
A simple power law can not fit all the points together:
there are at least three disctinct contributions (the
optically thick synchrotron emission from the jet, the thermal 
disk and Comptonization of soft photons by a hot medium).}
\end{figure}

We calculated for radio, NIR, optical
and X-ray data the corresponding flux value in $\nu F_\nu$ units
(erg~cm$^{-2}$~s$^{-1}$) corrected for extinction.  
Using our estimated value of $E_{\rm B-V}$ equals to 0.34$\pm$0.04~mag 
and the formula of
Predehl \& Schmitt (1995), we found $A_V \sim 3.1\times E_{\rm
B-V}=1.05\pm0.12$~mag. We then obtained
$A_\lambda$, respectively in the $B$, $R$ and $I$ filters (i.e. $\lambda$
equals 4400, 7100 and 9700~\AA) and in the REM filters with the formula of
Mathis \etal (1992) and Cardelli \etal (1989) depending on the wavelength. We 
got $A_B=1.54\pm0.18$, $A_R=0.68\pm0.08$, $A_I=0.40\pm0.05$,
$A_J=0.30\pm0.12$, $A_H=0.20\pm0.12$ and $A_K=0.12\pm0.12$~mag. Then, the
magnitudes were 
dereddened and we finally estimated the corresponding fluxes and
errors in units of erg~cm$^{-2}$~s$^{-1}$. {\it RXTE} and {\it INTEGRAL} data
were added to this plot which represents the 
Spectral Energy Distribution (SED) of \sw. We show our results from radio to 
X/$\gamma$-rays in Figure~\ref{sed} using a
logarithmic scale: the X-axis is labeled as log($\nu$) units and Y-axis in log
($\nu$ F$_\nu$).\\

From the analysis of the SED, we can confirm that the source shows all the
usual signs of a BH XN in the LHS. Firstly, we immediately note that the
source has a hard power law index in the high-energy domain. It is also clear
that a simple power law can not fit all the data together from radio to
X/$\gamma$-rays; at least two breaks are necessary in the SED. Secondly, the radio
emission is usually interpreted in this state as synchrotron emission
emanating from a self-absorbed compact jet (Blandford \& Konigl 1979). 
So in our case the flat radio power law
component could be explained by the optically thick synchrotron emission from
the jet (Markoff \etal 2005). From the SED we derive the power law index
$\alpha$ (in $S_\nu$): $\sim$ 0.03 in the radio (see 3.4) and $\sim$1.1 in the
optical/NIR. This latter value is compatible with the disk emission according
to the results reported by Hynes \etal (2005) on several X-ray binaries and
with the value observed (for example) in XTE~J1118$+$480 by Chaty \etal
(2003). If we extend the power law spectrum derived from the radio data points 
up to the NIR-optical frequencies with the same slope, the extrapolated 
flux is compatible with the first NIR points, in the $K$ band (see Figure~\ref{sed}),
but not with the rest of the measures. 
This suggests that the synchrotron emission from the compact jet 
of \sw~may contribute significantly to the $K$ band (as seen 
in certain BH in LHS, e.g., Russell et al. 2006 and references therein)
but little to the entire NIR-optical flux. This latter is probably dominated
by the disk emission.\\

The SED shown in Figure~\ref{sed} reveals that at least three distinct
contributions are necessary to account for our multiwavelength data (as a
single power law can not do so): the jet, the disk and the corona. The
necessity of an extra component to the jet to take into account our $B$, $V$,
$R$, $I$, $J$, $H$ and $K$ measured fluxes may represent part of the black
body disk contribution (thermal emission) instead of the (faint) companion. 
However, we do not have enough data in the required energy range (0.1--2~keV)
to constrain and modelize the SED: the disk contribution is faint in our data
range. We can only state that a break is necessary in the previous power law
(from radio) and that at least another component than the jet contributes in
the NIR-optical range. 
Then another break between the NIR-optical and soft X-ray data ($<$ 1
keV) is needed (in addition to the high-energy break around 50--100~keV of the
X/gamma-ray LHS spectrum) to model our X-ray/$\gamma$-ray data points of the
source from {\it RXTE} and {\it INTEGRAL}: these points have indeed a steeper
slope than the radio, IR and optical ones. The shape of the SED is similar to
the ones observed for transient LMXBs (e.g., XTE~J1118$+$480, Chaty \etal
2003; XTE~J1720$-$318, Chaty \& Bessolaz 2006).\\ 

\subsection{X and $\gamma$-ray constraints}

During the broad-band (3~keV--1~MeV) observations of \sw~presented in this
paper, the source was detected in a LHS well characterized by a thermal
Comptonization model modified by reflection. The best-fit parameters we obtain
through the {\it RXTE} and {\it INTEGRAL} data are consistent with those found
in BH binaries in the LHS \citep{McClintock03}. The Comptonization results
reported in this work are typical of a LHS as seen in the prototypical BH
Cygnus~X-1 (Cadolle Bel \etal 2006). Indeed, our reported results obtained on
Cygnus~X-1 during the LHS indicated a high Comptonization parameter
($\sim$0.51) like in the present case ($\sim$0.33). 
The reflection fraction we
derived is also compatible with values previously found for sources in the
LHS. The high peak luminosity, the fast rise, the slow decay time scales and
the bright hard state with spectral parameters and variability properties 
typically observed in other (dynamically confirmed) BH transients in LHS like
\eg~XTE~J1550$-$564 (Sobczak \etal 2000; Rodriguez \etal 2003), GRO~J1655$-$40
(Sobczak \etal 1999, see also McClintock \& Remillard 2006) or Nova Persei 92
(Roques \etal 1994, Denis \etal 1994, Finoguenov \etal 1996) clearly show that
\sw~is very likely a new XN and BHC.\\

The fact that we could not constrain the disk emission in LHS may be due to
its very low inner temperature which implies a negligible contribution at more
than 3~keV. This is compatible with the results obtained seven months later by
Miller \etal (2006) who reported that the disk component, well constrained by
the {\it XMM}/EPIC-pn spectrum, was very cold (0.22~keV). Besides, the lack of
strong Fe line is surprising with our reflection fraction 
value but a possible
interpretation is that the Fe line is originated from the outer disk and we do not
observe it due to inclination effect of the orbit.\\

While our data start at 3~keV, leading to a possible underestimation of the
bolometric luminosity, we derive an unabsorbed 2--11~keV flux of $1.5 \times
10^{-9}$~erg~cm$^{-2}$~s$^{-1}$, an unabsorbed 20--500~keV flux of $8.3 \times
10^{-9}$~erg~cm$^{-2}$~s$^{-1}$ and a bolometric flux (extrapolated from
0.01~keV to 10~MeV) of 1.3$ \times 10^{-8}$~erg~cm$^{-2}$~s$^{-1}$. This
corresponds to an unabsorbed bolometric luminosity of 5.77~$(d/6~{\rm
kpc})^2$$\times$~10$^{37}$~ergs~s$^{-1}$, well below the Eddington regime 
for a stellar mass BH. Miller \etal (2006) reported more
than 7 months after our observations that the unabsorbed flux in the
0.5--10~keV range was $3.9 \times 10^{-10}$~erg~cm$^{-2}$~s$^{-1}$ while
during our observations we get the higher value of $2.1 \times
10^{-9}$~erg~cm$^{-2}$~s$^{-1}$. Stellar mass BH accreting at or below
10$^{-2}$ $L_{\rm Edd}$ are found in the LHS (McClintock \& Remillard 2006).
These results are again 
compatible with the fact that the source was in the LHS during
the rise and the decline of the outburst.\\

\begin{figure}[t!]
\center
\resizebox{0.8\hsize}{!}{\includegraphics{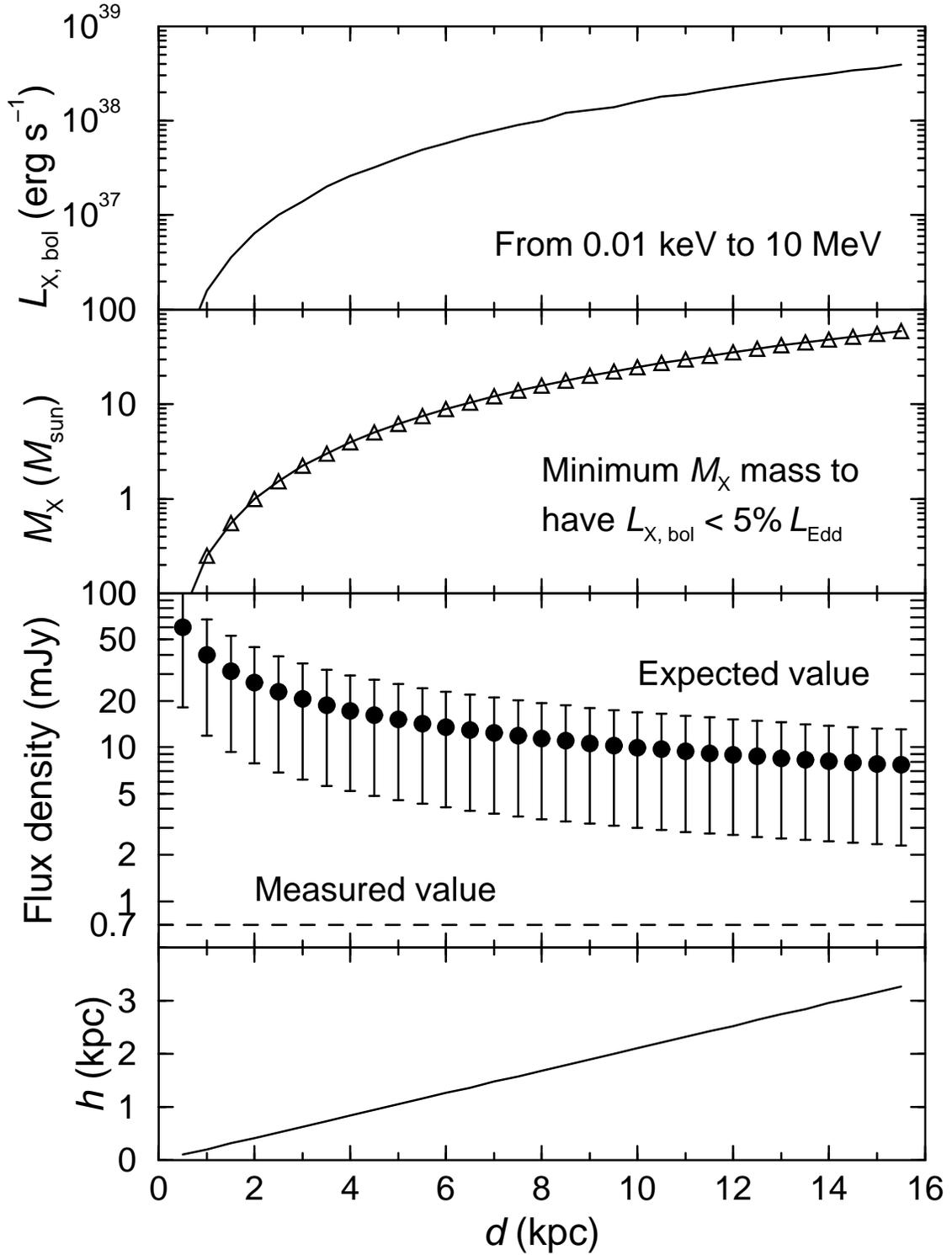}}
\caption{From top to bottom as a function of the distance:  bolometric
unabsorbed luminosity in the 0.01~keV--1~MeV band; 
minimum compact object mass to
guarantee $L_{\rm bol}$ $<$ 5 $\%$ $L_{\rm Edd}$; predicted radio flux
densities of SWIFT~J1753.5$-$0127 (filled circles with 1$\sigma$ error bars)
by using our measured X-ray flux and the correlation found for BH in the LHS
(Gallo \etal 2003) and our measured radio flux density at 8.5~GHz (dashed
line) showing a discrepancy of one order of magnitude even for large
distances; and height above the galactic plane.}
\label{corr}
\end{figure}

On the other hand, we can take the bolometric flux between 0.01~keV and
10~MeV, compute the bolometric luminosity for different distances to \sw~and
derive the minimum compact object mass to guarantee that this corresponds to
less than 5$\%$ of the Eddington luminosity, as seen in BH in the LHS
(Maccarone et al. 2003). These results are shown in the first two panels of
Figure~\ref{corr} and in Table~\ref{tabcorr} (for the minimum compact object
mass). If the overall 0.01~keV to 10~MeV spectrum had the same shape during
the maximum of the X-ray outburst, since the {\it RXTE}/ASM X-ray flux was a
factor of 3 higher than during our observations, the derived minimum compact
object masses should be a factor of 3 higher than those quoted in
Table~\ref{tabcorr}. However, since we do not have precise broad-band spectral
informations during the maximum of the outburst, we will use hereafter the bolometric
X-ray flux at the time of our observations, as well as the derived minimum
compact object masses for this flux.\\

Besides, the QPO frequency evolves from the beginning of the outburst to our
{\it INTEGRAL} observations: it decreases from the average value of 0.60
(Morgan \etal 2005) to 0.24~Hz. If the low frequency QPO is related to the
inner disk radius, this result would indicate that this radius increases 
during the decay of the outburst as seen before in, e.g.,
XTE~J1550$-$564 (Rodriguez \etal 2004a). This tendency is consistent with the
late stages of a LMXB outburst. These results are also compatible with the
ones reported by Miller \etal (2006) who found, seven months after our ToO, a
strong, band-limited variability noise (30$\%$ r.m.s. noise amplitude in the
different range 0.01--100~Hz) which is typical of the LHS in accreting BH,
while they do not detect any QPOs anymore. 
On the other hand, the QPO frequency
evolution may have another possible interpretation: it could be linked to
pulsational modes in the corona as recently suggested by Shaposhnikov \&
Titarchuk (2006).\\

\subsection{Optical constraints}

The blue optical spectrum seems to be the emission from an accretion disk
while the absence of strong emission lines points towards a late-type
companion star. If this is the case we should expect a fainter 
quiescent magnitude (consequently much 
higher than the $V=15.9\pm0.1$~mag value 
while in outburst reported by
Still \etal 2005). Usually, the optical magnitudes 
of BH XN change of about 5 mag from the 
peak of the outburst to the quiescence level as observed in other 
eruptions (e.g., GS~1354$-$64, Brocksopp \etal
2001). This trend is similar 
to the one shown by our results in spite of the fact that 
\sw~magnitudes increase more slowly. Besides, 
the source is not visible in archival images. Comparing with
nearby faint  USNO-B1.0 stars (Monet \etal 2003), we estimate a fainter 
quiescent visual magnitude (i.e. above 19.5~mag 
from $R > 19.0\pm0.5$ and $B > 20.0\pm0.5$).
Using for example the absolute visual magnitudes from Ruelas-Mayorga (1991),
even for the less luminous intermediate type giant companions in the range
F8-G2\,III, the distance to the source should be $\sim$15~kpc, implying a very
high minimum BH mass of $\sim$55~$M_\odot$ to guarantee $L_{\rm
bol}<5\%\,L_{\rm Edd}$. The situation would be much worse for other spectral
types. Clearly, an intrinsically fainter donor is required. To have a BH mass
below $\sim$10~$M_\odot$ (for $L_{\rm bol}<5\%\,L_{\rm Edd}$), the distance
has to be below 6~kpc, requiring for the companion star 
a spectral type later than G0\,V, for a
$V=19.5$ quiescent magnitude, or later than K0\,V, if the
quiescent magnitude is $V=21.0$ (i.e., $\sim$5 magnitudes fainter than the
$V=15.9\pm0.1$~mag value while in outburst). Therefore, the optical results
suggest a (main sequence) type K or M companion rather than earliest types,
ranging \sw~in the LMXB class. However, as the emission is dominated by the
disk and we could not detect any clear absorption lines, it is impossible to 
derive for sure the spectral type of the companion. 
In addition, there is no change in the $E_{\rm B-V}$ value.\\

The value of hydrogen column density we have derived thanks to the EW of the
NaI Doublet, $N_{\rm H}=(1.97 \pm 0.23)\times10^{21}$~atoms~cm$^{-2}$, is very
similar to the one measured with by {\it Swift}/XRT, of $N_{\rm
H}$~$\sim$~2.0~$\times$~10$^{21}$~atoms~cm$^{-2}$ 
(Morris et al. 2005). There is no need to 
advocate for extra intrinsic absorption around the X-ray
emitting source and a reasonable distance for this LMXB, probably located in
the Galactic Halo, could be 6~kpc. This corresponds to a galactic height of
1.3~kpc (see lower panel of Figure~\ref{corr}). 
This has to be compared with the
LMXB Galactic Halo sources XTE~J1118+480, 
located at 1.72~$\pm$~0.10~kpc (Gelino
et al. 2006) with a height above the Galactic Plane of 1.5~$\pm$~0.1~kpc, and
Scorpius~X-1, located at 2.8~$\pm$~0.3~kpc (Bradshaw et al. 1999) with a height
of 1.1~$\pm$~0.1~kpc. Therefore, a distance of 6~kpc to \sw~would place it at a
similar height above the Galactic Plane than those of XTE~J1118+480 and 
Scorpius~X-1.\\

\subsection{Radio constraints and discrepancy with the radio/X-ray correlation}

The point-like nature of the source during the MERLIN observations provides a
brightness temperature of $T_{\rm b}=1.0\times10^4$~K (see, e.g., Mart{\'i}
\etal 1998), thus practically ruling out thermal emission mechanisms. On the
other hand, the flat radio spectrum of \sw~is similar to the ones typically
found in BH during LHS (e.g., Fender \etal 2006 and references therein). This
is compatible with (and usually interpreted as) synchrotron radiation produced
in a partially self-absorbed conical and compact jet (Gallo \etal 2003, Fender
\etal 2004, Corbel \etal 2004). It is not resolved in our 
VLA data because we do not have enough angular resolution, contrary to the cases of Cygnus~X-1 or GRS~1915$+$105 (Stirling et al. 2001; 
Fuchs et al. 2003; Rib\'o 2005) observed with the VLBA. Even MERLIN 
could not resolve the source.\\

Corbel \etal (2003) then Gallo \etal (2003) found the following 
correlation between the
X-ray flux and the radio flux density for BH in the LHS: $F_{\rm rad} = k
F_{\rm X}^{+0.7}$ (with $k$=223~$\pm$~156 for all 
fluxes scaled to a distance of
1~kpc). We have used our measured unabsorbed X-ray flux ($F_{\rm
2-11~keV}=1.5\times10^{-9}$~erg~cm$^{-2}$~s$^{-1}$) to compute the expected
radio flux density according to their correlation by using different possible
distances to \sw. We proceed as follows: for each considered distance we
scaled our measured X-ray flux to the one we should detect at 1~kpc distance,
we then computed the expected radio flux at 1~kpc according to the
correlation and we finally derived the expected radio flux we should detect if
the source is placed at the considered distance. The obtained results are
quoted in the third column of Table~\ref{tabcorr} and are shown in the third
panel of Figure~\ref{corr}. It is clear from 
this Figure that the measured value
is one order of  magnitude lower than the expected one according to the
radio/X-ray correlation, even for the highest possible distances to the
source. We have performed the same analysis using the 
recent results of Gallo \etal 
(2006) who found a lower index of $\sim$0.6
in the radio/X-ray correlation: the same procedure, using this slightly 
flatter correlation (which only modifies significantly 
the lower X-ray flux part), provides very similar results. 
Indeed, since our source is approximately at the level of 
0.1 Crab (for a distance of 1 kpc) and at the level of 2 Crab 
(for a distance of 5 kpc), we are in the X-ray 
luminous part of the original Corbel 
et al. (2003) and Gallo et al. (2003) correlations: the new flatter 
correlation does not change significantly 
our results and, in any case, they are well within the 
errors of those obtained with the original 
correlation in this part of the diagram.\\

For comparison, the MERLIN observation performed on MJD~53555
(centered on 2005 July 3 at 23:00) took place when the {\it RXTE}/ASM X-ray
flux was higher with a factor of 2.9 (13~$\pm$~1 compared to
4.5~$\pm$~0.5~count~s$^{-1}$). From this we can predict a factor of 2.15 higher
for the radio emisssion while the radio flux density was 3 times higher.
Therefore, similar (although slightly smaller) discrepancies are obtained in
this case. Therefore, this would point to a different track in the 
radio/X-ray fluxes correlation diagram than the one previously observed 
(Gallo et al. 2003). This could indicate that the compact jet formation 
and luminosity may be able to evolve differently than previously thought 
(see also Corbel et al. 2004 for XTE J1650$-$500; Rodriguez et al. 2006b for
IGR~J17497$-$2821 and Chaty 2006 for XTE~J1720$-$318).\\ 

\section{Conclusions}

In recent years, it has become apparent that in the LHS, the BH binaries
become bright in radio (Fender \etal 2006 for a complete review) and display
clear correlations between the X-ray and radio luminosities
\citep{corb03,gallo03,gallo06,russ06} as observed, for example, 
in the confirmed BH system 
Cygnus~X-1 \citep{brock99,gleiss04,nowak05}. Models where 
the base of a compact jet plays
a major role in the physical processes of such BH systems have been proposed
(Markoff \& Nowak 2004; Markoff \etal 2005): 
these authors have computed jet models
where the synchro-self Compton or the external Comptonization radiation are
the dominant processes generating X-ray spectra in BH binaries. 
The high-energy emission seen during the LHS is interpreted as
synchrotron, Comptonization and reflection emission from the jet that extends
from radio to hard X-rays, naturally explaining the correlations observed
during the LHS.\\

We have accurately studied \sw~and found that, although clearly in LHS, this
source is interestingly well below the radio/X-ray correlation 
laws of BHs binaries in LHS determined previously by different authors 
(Corbel et al. 2003; Gallo et al. 2003, 2006), 
even assuming a large distance. 
Another possibility is that \sw~could radiate at a 
luminosity higher than 5$\%$ of the
Eddington regime (while it stayed in the LHS). 
In our case, thermal Comptonization models fit relatively well our
simultaneous \emph{RXTE}/\emph{INTEGRAL} data. However, the source and
geometry of Comptonization (torus around the BH or corona over the disk) are
not defined. The emission could also be compatible with the standard picture
of synchrotron and inverse Compton radiation coming from a self-absorbed
conical jet (Markoff et al. 2005). Its 
contribution in the IR and optical ranges,
which is probably dominated in our case by the emission from a cold thermal
disk, is small.\\

Thanks to our {\it INTEGRAL} AO3 proposal triggered simultaneously 
with {\it
RXTE}, NTT, REM and VLA, it has been possible to perform
multiwavelength observations of the XN \sw~in the LHS on a wide energy band.
With the combination of all high-energy (3~keV--1~MeV) instruments on board
{\it RXTE} and {\it INTEGRAL}, we have measured crucial spectral and timing
parameters on \sw~during its hard outburst. This gives clues on the BH nature
of this object. \sw~could belong to a small subset of BHCs that are
only observed in the LHS while in outburst. This subset includes the
well-known source XTE~J1118$+$480 (e.g., Hynes \etal 2000, 2003). The
classification in the LMXB is obtained from our optical results. Radio data
confirm the presence of a partially self-absorbed conical jet 
and indicate that its radio luminosity is an order of magnitude lower 
than the one previously typically observed for a BH in the LHS. 
Our results imply that the source 
could be located at the reasonable distance of $\sim$6~kpc.\\

The detection and study of a large sample of XN in outburst (and in decay
phase) will provide accumulated multiwavelength data on bright sources in
order to obtain crucial evolutions of the spectra, therefore improving our
understanding of the galactic BH physics and origin. Thanks to large changes
in a short period of time (as usually the case for LMXB), it will thus shed
light on the physics, accretion processes and radiation mechanisms at work in
the vicinity of BH binaries.\\

\acknowledgments

We thank the {\it INTEGRAL} and {\it RXTE} mission 
planners for programming
the ToO observations described in the paper.
M.R. acknowledges financial support from the French Space Agency (CNES) and
from the Spanish Ministerio de Educaci\'on y Ciencia through a \emph{Juan de
la Cierva} fellowship linked to the project AYA2004-07171-C02-01, which is
partially supported by FEDER funds.
SC thanks the ESO staff and especially C\'edric Foellmi for performing these
ToO observations. 
PU and AT has been supported by the Italian Space Agency via the grant
I/R/046/04. 
The present work is based on ESO observations (through part of program $\#$
075.D-0634) and with the use of {\it INTEGRAL}, an ESA project with
instruments and science data center funded by ESA member states (especially
the PI countries: Denmark, France, Germany, Italy, Switzerland, Spain, Czech
Republic and Poland, and with the participation of Russia and the USA).
This research has made use of the NASA Astrophysics Data System Abstract
Service and of the SIMBAD database, operated at the CDS, Strasbourg, France.

\clearpage

\begin{table*}
\begin{center}
\caption{Log of the \sw~observations analyzed in this paper.\label{log}}
\begin{tabular}[h]{llllll}
\\
\hline
Spacecraft & Instrument & Observation Period & Exposure & Observation \\
 & & (MJD-53590; U.T.) & (ks) & Type/Mode \\
\hline
{\it INTEGRAL}& JEM-X & 2.52--4.64; 12:29--15:31~$^a$ & 170 & ToO, HEX~$^b$\\
{\it INTEGRAL}& IBIS & 2.52--4.64; 12:29--15:31~$^a$ & 176 & ToO, HEX~$^b$\\
{\it INTEGRAL}& SPI  & 2.52--4.64; 12:29--15:31~$^a$ & 134 & ToO, HEX~$^b$\\
{\it Rossi-XTE} & PCA & 3.20; 4:51~$^c$& 3.1 & Public \\
{\it Rossi-XTE}& HEXTE & 3.20; 4:51~$^c$ & 2.2 & Public \\
~~~~~~- & NTT/EMMI & 3.08~$^c$; 02:07--05:42 & 3.6 & ToO \\
~~~~~~- & REM/ROSS~$^d$ & 3.98; 11.98~$^e$ & 1 & ToO \\
~~~~~~- & REM/REMIR & 3.03; 11.03~$^e$ & 7.3 & ToO \\
~~~~~~- & VLA & 3.19--3.35; 04:20--08:50 & 16.2 & ToO \\
\hline
\end{tabular}
\end{center}
Notes:\\ a) Plus 2 days;\\
b) Hexagonal pattern around the nominal target location;\\
c) At the beginning of the observation;\\
d) 0.6~ks observations also taken on 53599.99 (U.T. 17.99) 
and 53640.04 MJD (U.T. 27.04) respectively (see Section 3.5);\\
e) At mid-observations.\\
\end{table*}

\clearpage

\begin{table*}
\begin{center}
\caption{Best-fit parameters of individual components used to
fit the Power Density Spectrum (errors at 90$\%$ confidence 
level: $\Delta$\ki~=~2.7).\label{PDS}}
\begin{tabular}[h]{llll}
\\
\hline
Lorentzian \# & Centroid Frequency & Width & RMS amplitude \\
 & (Hz)& (Hz) & ($\%$) \\
\hline
n1 & 0 (Frozen) & 0.92$_{-0.06}^{+0.05}$ & 26.8$_{-1.9}^{+1.6}$ \\
n2 & 0 (Frozen) & 9.9$_{-1.3}^{+1.4}$ & 16.8$_{-2.6}^{+2.9}$ \\
QPO & 0.241$\pm$0.006 & 0.03$_{-0.01}^{+0.02}$ & $5.4_{-1.9}^{+1.7}$\\
\hline
\end{tabular}
\end{center}
\end{table*}

\clearpage

\begin{table*}
\begin{center}
\caption{Summary of radio observations.
\label{tabradiospec}}
\begin{tabular}[h]{lllll}
\\
\hline
$S_{\rm 1.4~GHz}$ & $S_{\rm 4.9~GHz}$ & $S_{\rm 8.5~GHz}$ & $S_{\rm 15~GHz}$ & \\
(mJy) & (mJy) & (mJy) & (mJy) & $\alpha_{\rm \,1.4-8.5~GHz}$\\
\hline
$0.65\pm0.15$ & $0.65\pm0.07$ & $0.7\pm0.1$ & $<$0.60 & $+0.03\pm0.03$ \\
~~~~~~~~$(0.04)$ & ~~~~~~~~$(0.04)$ & ~~~~~~~$(0.01)$ & ~~~~~$(0.13)$ \\
\hline
\end{tabular}
\end{center}
Note: Errors are at the 1$\sigma$ level and a 3$\sigma$ upper limit 
is reported at 15~GHz. The expected r.m.s. sensitivities are given in parentheses.
\end{table*}

\clearpage

\begin{table*}
\begin{center}
\caption{Minimum compact object mass to guarantee $L_{\rm bol}$
$<$ 5$\%$ $L_{\rm Edd}$ and expected radio flux densities for several
distances to \sw~(according to the relation of 
Gallo et al. 2003, see text).
\label{tabcorr}}
\begin{tabular}[h]{lll}
\\
\hline
$d$ & Minimum compact & ~~~$S_{\rm 5~GHz}$ \\
(kpc) &  object mass (M$_\odot$) & ~~~(mJy)\\
\hline
  1  &  ~~~~~~ 0.3  &  $39.71\pm27.78$  \\
  2  &  ~~~~~~ 1.0  &  $26.20\pm18.33$  \\
  3  &  ~~~~~~ 2.2  &  $20.54\pm14.37$  \\
  4  &  ~~~~~~ 4.0  &  $17.28\pm12.09$  \\
  5  &  ~~~~~~ 6.2  &  $15.12\pm10.58$  \\
  6  &  ~~~~~~ 8.9  &  $13.55\pm 9.48$  \\
  7  &  ~~~~~~ 12.1 &  $12.35\pm 8.64$  \\
  8  &  ~~~~~~ 15.8 &  $11.40\pm 7.98$  \\
  9  &  ~~~~~~ 20.0 &  $10.63\pm 7.43$  \\
 10  &  ~~~~~~ 24.7 &  $ 9.97\pm 6.98$  \\
 11  &  ~~~~~~ 29.9 &  $ 9.42\pm 6.59$  \\
 12  &  ~~~~~~ 35.5 &  $ 8.94\pm 6.25$  \\
 13  &  ~~~~~~ 41.7 &  $ 8.52\pm 5.96$  \\
 14  &  ~~~~~~ 48.4 &  $ 8.15\pm 5.70$  \\
 15  &  ~~~~~~ 55.5 &  $ 7.82\pm 5.47$  \\
\hline
\end{tabular}
\end{center}
Note: Errors at the 1$\sigma$ level.
\end{table*}

\clearpage

\clearpage


\clearpage


\clearpage


\clearpage


\clearpage


\clearpage

\clearpage


\clearpage

\clearpage


\clearpage

\clearpage


\end{document}